HE.5.1.02

# Detection of Galactic Dark Matter by GLAST

**Alexander Moiseev[1,2], Jonathan Ormes[1], Heather Arrighi[1], Elliott Bloom[3], Chris Chaput[3], Seth Digel[1], Daniel Engovatov[3], Jay Norris[1], and Jeff Silvis[1]**
[1]NASA/Goddard Space Flight Center, Greenbelt, MD 20771, USA
[2]University Space Research Association, Seabrook, MD 20706, USA
[3]Stanford Linear Accelerator Center, Stanford, CA 94309, USA

**Abstract**
The mysterious dark matter has been a subject of special interest to high energy physicists, astrophysicists and cosmologists for many years. According to theoretical models, it can make up a significant fraction of the mass of the Universe. One possible form of galactic dark matter, Weakly Interacting Massive Particles (WIMPs), could be detected by their annihilation into monoenergetic gamma-ray line(s). This paper will demonstrate that the Gamma-ray Large Area Space Telescope (GLAST), scheduled for launch in 2005 by NASA, will be capable of searching for these gamma-ray lines in the energy range from 20 GeV to ~500 GeV and will be sufficiently sensitive to test a number of models. The required instrument performance and its capability to reject backgrounds to the required levels are explicitly discussed.

## 1. Introduction

On many scales, from galaxies to the largest structures in the Universe, there is a discrepancy between observed (luminous) matter in the Universe and that inferred from dynamical considerations. This is seen on the scale of our own galaxy. Galactic dark matter was suggested to solve this discrepancy (Trimble, 1987; Sikivie, 1995); one possible form could be the proposed, but undiscovered, SUSY particles known as WIMPs (Weakly Interacting Massive Particles). WIMPs can be detected through stable products of their annihilations: energetic neutrinos, antiprotons, positrons, gamma-quanta etc. (Jungman, Kamionkowski & Griest, 1996; and references therein). It is very important to search for a signature of WIMPS which could not be misinterpreted. In principle, WIMPs cannot annihilate directly to photons, but there should be a small cross section into photons through intermediate one-loop processes. In this case there should be high energy (10-1000 GeV) monochromatic gamma-lines; the lines should be very narrow because of the low velocity of WIMPs in the galaxy. Estimations of the possible intensity of these lines depend upon a number of assumptions. The highest gamma-line intensities are predicted assuming WIMPs have condensed into the Galactic Center or into clumps in the galactic plane. For example, Bergstrom, Ulio and Buckley (1998) show that some models might produce a flux as large as $\sim 2\times 10^{-11}$ cm$^{-2}$s$^{-1}$ at 100 GeV from the $10^{-5}$ sr cone around the Galactic Center.

## 2. Conditions of the experiment

The requirements for an experiment to search for possible lines are that the lines should be seen above a background which is a continuum of galactic gamma rays. Optimally there should be negligible residual contamination from cosmic rays misidentified in the detector. Thus, energy resolution and geometry factor/sensitive area of the detector, background rejection, and the observation time are the critical factors to be optimized.

GLAST is a mission scheduled for launch in 2005 to continue the detailed exploration of the Universe in >100 MeV gamma-rays began by EGRET (Atwood et al., 1994). GLAST is sensitive to gamma radiation in the range of 30 – 300 GeV (fig.1). The energy of the detected photon is measured by a CsI calorimeter, which is situated below the tracker, and has a size of 170 cm × 170 cm × 20 cm. The effective calorimeter thickness for the normally incident particles is ~10 radiation lengths which provides ~10% energy resolution at 300 GeV.

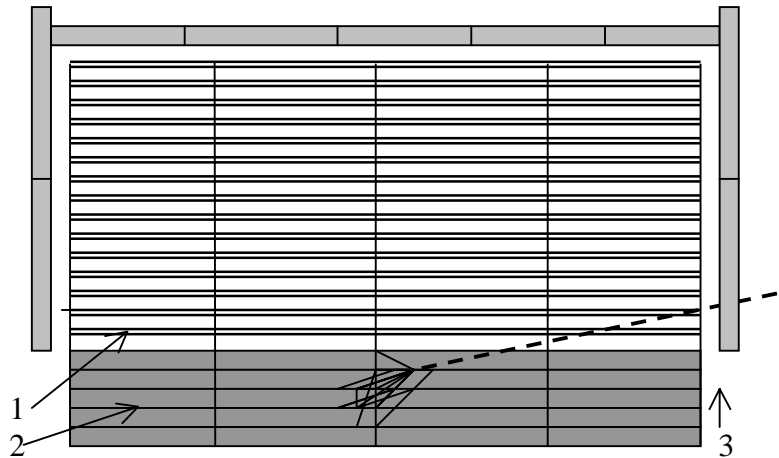

Fig.1. Simplified schematic view of GLAST (number of tracker planes and calorimeter layers is reduced). 1 – tracker, 2 – calorimeter, 3 – ACD.

Much better energy resolution can be achieved for the off-angle events with longer paths in the calorimeter (shown in fig.1).

**2.1. Background rejection**. The first task of GLAST is to remove the abundant background of cosmic ray protons and helium nuclei whose differential flux is 5 orders of magnitude higher than that of high latitude diffuse gamma radiation at 30 GeV. We also must consider cosmic ray electrons, which are 1000 times more abundant. In order to carry out a sensitive search for gamma ray lines one should be able to reject protons (electrons) with power better than $3\times10^6$ and $3\times10^4$ respectively. The main strategy for proton rejection is the following: a track image in the tracker and lateral and longitudinal profile of the shower in the calorimeter provide at least $10^3$ of the rejection (Norris et al., 1997; Ormes et al., 1997), and an anticoincidence detector (ACD) provides remaining $3\times10^3$ (Moiseev et al. 1999). The cosmic ray electrons are the more serious enemies. Their showers are identical to those of photons in the calorimeter since both are electromagnetic. Thus, the ACD is the main defense against electrons and should have rejection power to minimum ionizing charged particles of ~3000. An additional factor of 10 comes from the use of the tracker to reach the $3\times10^4$ requirement. We have measured the rejection power (efficiency) of the scintillator paddles we plan to use for GLAST and find there are sufficient photo-electrons to obtain $> 3\times10^3$ rejection for a threshold setting $< 0.3 \times$ mip.

To find the ACD tile which was crossed by a detected off-angle particle, which has a longer path in a calorimeter, we follow a two step process. First we use the imaging capability of the calorimeter and reconstruct a trajectory with a precision of 2-3 degrees (Norris et al., 1997). Then we project that cone back into the tracker and look for the absence of a track (for photons) in the past two layers before the ACD and the absence of a hit in the ACD tile for electrons; the two tracker hits (for electron) provide precise pointing to an ACD tile.

**2.2. Backsplash.** Use of an ACD creates the problem of backsplash. High energy electromagnetic cascades produce soft radiation, mainly minimum attenuation photons, that escape from the calorimeter. These photons can produce a Compton electron in an ACD and create a self-veto, making the instrument insensitive to gamma-radiation above 50-100 GeV. For EGRET, built with a monolithic ACD dome, this effect reduced the efficiency by a factor of 2 at 30 GeV (Thompson et al.,1993). To minimize the effect of backsplash, GLAST has a segmented ACD, and only the tile crossed by the projected event trajectory is used for vetoing an event. The required ACD segmentation was studied in detail both using Monte Carlo simulations and in SLAC beam test (Atwood et al., 1999; Moiseev et al., 1999). On the top of GLAST, the ACD segmentation of ~1000 cm$^2$ is sufficient to maintain >90% efficiency to the highest energies. We wish to use events that enter GLAST at $>60^0$ incidence angle to obtain a sample with few percent energy resolution. These events enter through the sides of GLAST. For them the calorimeter, and consequently the source of backsplash, are closer to the ACD. The closer the ACD tiles to the "source" of backsplash, the smaller the tiles must be. From our SLAC beam test and Monte Carlo study we have shown that the backsplash into a given solid angle is almost uniform within a $60^0$ cone in the backward direction (Atwood et al., 1999). The required segmentation can be given by $A_{90\%} = (R/60)^2 \times 1000$ cm$^2$ where R[cm] is the distance from the ACD tile to the calorimeter. We find that on the side of GLAST a tile size of ~200 cm$^2$ limits backsplash caused self-veto to be less than 10% at 300 GeV.

## 3. Capability of GLAST to detect gamma lines

Energy resolution for 50-500 GeV photons should reach several percent for path lengths of >20 $X_0$. The effective area for the GLAST calorimeter for such trajectories is estimated to be $2 \times 10^4$ cm$^2$. The geometry factor of GLAST for isotropic flux through the top and sides of the tracker is shown in fig.2 where the Earth obscuration is accounted for by a factor of 0.74 (assuming zenith pointing) applied to the events entering through the tracker sides. To take into account the requirement that the trajectory passes through at least two tracker trays, the geometry factor given in fig.2 is calculated for events which cross ACD at least 6 cm above the top of calorimeter.

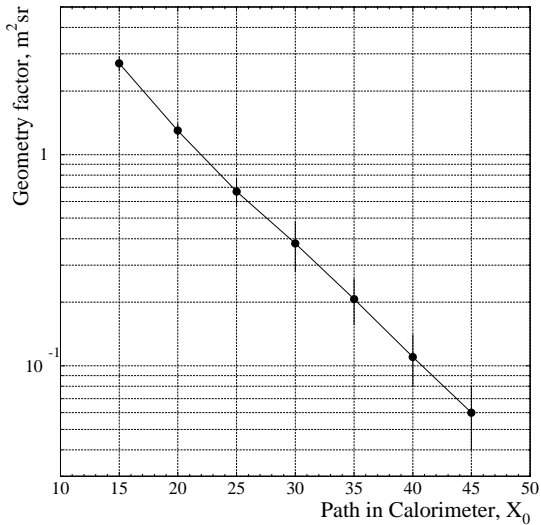

Fig.2. Geometry factor for the GLAST calorimeter

The capability of GLAST's calorimeter to detect photons with high energy resolution was simulated with the event generator Glastsim. A set of event cuts was developed to select events with shower containment in the calorimeter that provides the best energy resolution. The initial selections were optimized to achieve the best energy resolution while maximizing the fraction of retained photons. The geometry factor strongly increases when we accept events with shorter path lengths, so the minimization of the required pathlength was important. The results of simulations can be summarized as follows: **2-3% energy resolution** is achievable while retaining **~50% of the photons** and requiring the pathlength to be **more than 20$X_0$**. The gain variation from CsI crystal to crystal was assumed to be within 1%, a challenging problem for a long duration space experiment.

## 4. Summary

Here we present the sensitivity of GLAST based on two possible models for the distribution of WIMPs in the galaxy. One model is a "dark matter point source" in the Galactic Center assuming the WIMPs have fallen into a small region within $10^{-5}$ sr (Bergstrom, Ullio and Buckley, 1998). The second model assumes a broad distribution falling off like high latitude diffuse radiation.

For high latitude model the sensitivity $I_\gamma$ of GLAST for a gamma-line of energy $E_\gamma$ approximately is given by

$$I_\gamma = \frac{n_\sigma}{0.68} \sqrt{\frac{2 F_b \eta E_\gamma}{GT}}$$

and for Galactic Center model

$$I_\gamma = \frac{n_\sigma}{0.68\sqrt{S f_t T}} \sqrt{2\eta E_\gamma (F_{GC} + F_b \Delta\Omega)}$$

where $n_\sigma$ is the significance (in $\sigma$), $F_b$ is the background flux, $F_{GC}$ is the differential gamma-radiation from the Galactic Center, G is the instrument geometrical factor, S is the sensitive area, $\eta$ is the relative energy resolution (half width containing 68% of events), T is the observation time, $2\eta E_\gamma$ is the binning width, $\Delta\Omega = 10^{-3}$ sr is the point-spread function for the calorimeter, and $f_t$ (0.25) is the fraction of time during which the Galactic Center lies in a direction that provides a path length in a calorimeter of more than 20 $X_0$.

Table 1

| Energy of the line $I_\gamma$ | High Latitude Model Source [cm$^2$ s sr]$^{-1}$ | Galactic Center Model Source [cm$^2$ s]$^{-1}$ |
|---|---|---|
| 50 GeV | $1.8 \times 10^{-10}$ | $1.2 \times 10^{-10}$ |
| 100 GeV | $1.2 \times 10^{-10}$ | $8 \times 10^{-11}$ |
| 500 GeV | $5 \times 10^{-11}$ | $3 \times 10^{-11}$ |

Table 1 contains our estimates of the sensitivity to WIMP lines in GLAST for the case of the high latitude model, and the Galactic Center. To set the scale of sensitivity, we have arbitrary calculated $I_\gamma$ for a 3$\sigma$ signal. The GLAST observational parameters used were $\eta=0.02$, G=0.5 m$^2$ sr (efficiency of the event selection is taken into account), S=6000 cm$^2$ (effective area of the calorimeter to provide path length more than 20 $X_0$), and T = 3 years. More exact treatment raises the estimated sensitivity by a factor of 1.1-1.6. We used the high latitude gamma-radiation flux given in Sreekumar et al., 1998 and Galactic Center radiation from Hunter et al., 1997; Mayer-Hasselwander et al., 1998. The upper energy limit for this gamma ray line search is limited by the dynamic range of the calorimeter readout electronics; also at higher energies the low photon flux is the limiting factor. We note that our sensitivity is a few times higher than the optimistic estimates for the predicted flux (Bergstrom, Ullio & Buckley, 1998) but both the number of assumptions in the expected flux calculations and inestimable importance of a positive result motivated this work.

Acknowledgements. The authors are grateful to David Bertsch, Robert Hartman and David Thompson for their valuable comments.